\documentclass[12pt,a4paper,english,aps,pre,preprint,floatfix]{revtex4}
\usepackage[T1]{fontenc}
\usepackage[latin1]{inputenc}
\usepackage{graphicx}

\makeatletter

\usepackage{babel}
\makeatother
\begin{document}

\title{Stationary solutions and Neumann boundary conditions in the Sivashinsky
equation}

\author{Bruno Denet}

\affiliation{IRPHE 49 rue Joliot Curie BP 146 Technopole de Chateau Gombert 13384
Marseille Cedex 13 France}

\email{bruno.denet@irphe.univ-mrs.fr}


\preprint{submitted to Phys. Rev. E}

\begin{abstract}
New stationary solutions of the (Michelson) Sivashinsky equation of
premixed flames are obtained numerically in this paper. Some of these
solutions, of the bicoalescent type recently described by Guidi and
Marchetti, are stable with Neumann boundary conditions. With these
boundary conditions, the time evolution of the Sivashinsky equation
in the presence of a moderate white noise is controlled by jumps between
stationary solutions.
\end{abstract}
\maketitle

\section{Introduction \label{sec:Introduction}}

The Sivashinsky equation \cite{siva77} (or Michelson Sivashinsky
equation depending on the authors) is a well established non linear
equation which provides a satisfactory description of the time evolution
of premixed flames. Until working on the present paper, the author
had a very simple idea of the situation regarding this equation. Pole
solutions of the Sivashinsky equation were obtained in \cite{leechen}
and popularized in \cite{thualfrischhenon}, which reduces the time
evolution of the equation to a dynamical system and the stationary
solutions to finding zeroes of a non linear function of several variables.
The paper \cite{thualfrischhenon} also shows that the poles have
a tendency to coalesce, i.e. to align vertically in the complex plane.
Stationary solutions were obtained in the form of coalescent solutions
with a number of poles depending on the width of the domain. It was
shown analytically in \cite{vaynblattmatalon1}\cite{vaynblattmatalon2}
that each solution, with a given number of poles is linearly stable
in a given interval for the control parameter (either the domain width
or more often the curvature term with a domain width fixed to $2\pi$).
Numerical simulations however, always performed with periodic boundary
conditions, continue to show that the solutions are extremely sensitive
to noise \cite{olamiprocacciapre} for sufficiently large domains.
These results are consistent with a qualitative description of the
stability of curved flame fronts by Zeldovich et. al. \cite{zeldovichcurved}. 

For some reason, the author of the article began simulations of the
Sivashinsky equation with Neumann boundary conditions, ie. zero slope
of the flame front at each end of the domain. Of course, Neumann boundary
conditions are a more realistic description of a flame in a tube than
periodic boundary conditions. However, as solutions with Neumann boundary
conditions on $[0,\pi]$ are simply symmetric solutions with periodic
boundary conditions on $[0,2\pi]$, the author was thinking that he
should obtain basically a coalescent solution, but only between $0$
and $\pi$, with all the poles coalescing at $0$, leading to a cusp
at this boundary. It was so obvious that actually simulations of the
Sivashinsky equation with Neumann boundary conditions were only used
originally as a test case for a new computer program. However stationary
solutions were obtained, where poles did not all coalesce at the same
position, but actually on the two boundaries.

It turns out (although the author was absolutely unaware of this paper
at the beginning of his work) that this type of stationary solutions,
called bicoalescent solutions, were already discovered by Guidi and
Marchetti \cite{guidimarchetti}. In Section \ref{sec:Stable-bicoalescent-solutions},
we show the new bicoalescent solutions that we have obtained, which
have a nice property with Neumann boundary conditions, they are stable.
These solutions were not found in \cite{guidimarchetti} because the
curvature parameters studied were too large (or equivalently, the
domain width was too small). In Section \ref{sec:web-of-stationary},
we show where the new solutions of Section \ref{sec:Stable-bicoalescent-solutions}
are found in the parameter space. We have thus to study a larger domain
of the parameter space than in \cite{guidimarchetti}, and discover
also new stationary solutions of the interpolating type described
by Guidi and Marchetti (see section \ref{sec:web-of-stationary} for
a definition of this type of solution). These interpolating solutions,
unlike those of Section \ref{sec:Stable-bicoalescent-solutions},
are unstable. The number of stationary solutions obtained is so large
that we have entitled Section \ref{sec:web-of-stationary} web of
stationary solutions and will try to convince the reader that this
is not an exageration. In Section \ref{sec:evolution-with-noise},
the evolution of the Sivashinsky equation with noise is studied. In
the case of Neumann boundary conditions, as expected, the stable bicoalescent
solutions play a dominating role in the dynamics. Finally, Section
\ref{sec:conclusion} contains a conclusion.

\section{Stable bicoalescent solutions \label{sec:Stable-bicoalescent-solutions}}

The Sivashinsky equation can be written as 

\begin{equation}
\phi_{t}+\frac{1}{2}\phi_{x}^{2}=\nu\phi_{xx}+I\left(\phi\right)\label{eq:sivashinsky}\end{equation}

where $\phi\left(x\right)$ is the vertical position of the front.
The Landau operator $I\left(\phi\right)$ corresponds to a multiplication
by $\left|k\right|$ in Fourier space, where $k$ is the wavevector,
and physically to the destabilizing influence of gas expansion on
the flame front (known as the Darrieus-Landau instability). $\nu$
is the only parameter of the equation and controls the stabilizing
influence of curvature. The linear dispersion relation giving the
growth rate $\sigma$ versus the wavevector is, including the two
effects: 

\begin{equation}
\sigma=\left|k\right|-\nu k^{2}\label{eq:dispersion}\end{equation}

As usual with Sivashinsky-type equations, the only non linear term
added to the equation is $\frac{1}{2}\phi_{x}^{2}$. In the flame
front case, this term is purely geometrical : the flame propagates
in the direction of its normal, a projection on the vertical ($y$)
direction gives the factor $\cos\left(\theta\right)=1/\sqrt{1+\phi_{x}^{2}}$,
where $\theta$ is the angle between the normal and the vertical direction,
then a development valid for small slopes of the front leads to the
term $\frac{1}{2}\phi_{x}^{2}$. The Sivashinsky equation will be
solved numerically on $[0,2\pi]$ with periodic boundary conditions,
or (more often in this paper) on $[0,2\pi]$ with only symmetric modes,
which corresponds to homogeneous Neumann boundary conditions on $[0,\pi]$
(zero slope on both ends of the domain). All dynamical calculations
will be performed by Fourier pseudo-spectral methods (i.e. the non
linear term is calculated in physical space and not by a convolution
product in Fourier space). The method used is first order in time
and semi-implicit (implicit on the linear terms of the equation, explicit
on $\frac{1}{2}\phi_{x}^{2}$). No particular treatment of aliasing
errors has been used.

Pole solutions (\cite{thualfrischhenon}) of the Sivashinsky equation
are solutions of the form:

\begin{equation}
\phi=2\nu\sum_{n=1}^{N}\left\{ \ln\left(\sin\left(\frac{x-z_{n}(t)}{2}\right)\right)+\ln\left(\sin\left(\frac{x-z_{n}^{*}(t)}{2}\right)\right)\right\} \label{eq:poledecomposition}\end{equation}

where $N$ is the number of poles $z_{n}(t)$ in the complex plane.
Actually the poles appear in complex conjugate pairs, and the asterisk
in Equation \ref{eq:poledecomposition} denotes the complex conjugate.
In all the paper, only poles with a positive imaginary part will be
shown, the number of poles will also mean number of poles with a positive
imaginary part. The pole decomposition transforms the solution of
the Sivashinsky equation into the solution of a dynamical system for
the locations of the poles. In the case of stationary solutions, the
locations of the poles are obtained by solving a non linear system:

\begin{equation}
-\nu\sum_{l=1,l\ne n}^{2N}\cot\left(\frac{z_{n}-z_{l}}{2}\right)-i\textnormal{sgn}\left[{\textstyle \textnormal{Im}}\left(z_{n}\right)\right]=0\;\;\;\;\; n=1,\cdots,N\label{eq:nonlinearsystem}\end{equation}

where $\textnormal{Im}\left(z_{n}\right)$ denotes the imaginary part
and sgn is the signum function. This non linear system will be solved
by a Newton-Raphson method.

Let us define here a process that will be called folding in the rest
of the paper and which allows to create cellular solutions starting
from curved flame fronts (i.e. fronts with only one cell in $[0,2\pi]$).
If a solution $\phi_{1}\left(x\right)$ of the Sivashinsky equation
exists with parameter $1/\nu_{1}$, then $\phi_{2}\left(x\right)=\frac{1}{m}\phi_{1}\left(mx\right)$
is a solution of the Sivashinsky equation with parameter $1/\nu_{2}=m\left(1/\nu_{1}\right)$,with
$m$ integer. 

Although we have searched for stationary solutions with periodic boundary
conditions, it appears that all the solutions we have found on $\left[0,2\pi\right]$
are symmetric, and thus are stationary solutions with Neumann boundary
conditions, i.e. zero slope, on $\left[0,\pi\right]$. In most of
the cases the stationary solutions obtained have poles at $x=0$,
in a few cases however, the solutions have no poles on the boundaries
(i.e. only lead to symmetric solutions with no poles at the boundary)

With periodic boundary condition, the well-known result is that in
the window $2n-1$$\leq1/\nu$$\leq2n+1$, $n=1,2,\cdots$ there exists
$n$ different monocoalescent stationary solutions (all the poles
have the same real part), with $1$ to $n$ poles, and the solution
with the maximum number of poles $n$ is asymptotically stable. For
a particular value of $1/\nu$, the number $n(\nu)$ such that $2n-1$$\leq1/\nu$$\leq2n+1$
will be called the optimal number of poles. All stable solutions found
in this paper, for any value of $1/\nu$, even with Neumann boundary
conditions, have the optimal number of poles $n(\nu)$.

Using however the Sivashinsky equation (Eq. \ref{eq:sivashinsky})
with Neumann boundary conditions, we obtain in each of the intervals
$\left[2n-1,2n+1\right]$ of the parameter $1/\nu$, not only one
asymptotically stable solution, but several, of the form $\left(l,n-l\right)$
with $l=0,1,$$\cdots,n$ where $l$ poles coalesce at $x=0$ and
$l-n$ coalesce at $x=\pi$ (The bicoalescent type of solutions have
been recently introduced in \cite{guidimarchetti}). These solutions
will also be obtained from the non linear system of equations (Eq.
\ref{eq:nonlinearsystem}) in Section \ref{sec:web-of-stationary}.
It must be remarked that all these solutions, except the monocoalescent
one, are unstable for periodic boundary conditions, i.e. when antisymmetric
perturbations are allowed on $\left[0,2\pi\right]$. We have just
defined here the notation $\left(n_{1},n_{2}\right)$ that will be
used in the paper for bicoalescent solutions with $n_{1}$ poles at
zero, and $n_{2}$ at $\pi$. Monocoalescent solutions can be seen
as a particular case of bicoalescent solutions and will be noted $\left(n,0\right)$.
We will encounter also multicoalescent solutions, such as $\left(n_{1},n_{2},n_{3},\cdots\right)$,
which means that in the interval $\left[0,2\pi\right]$, the poles
coalesce at different locations: $n_{1}$ poles coalesce at a position
on the left of the interval, generally $0$, $n_{2}$ poles coalesce
at a position with a higher value of $x$, then $n_{3}$ at a position
with a value of $x$ even higher, and so on. With this notation (1,1,1)
represents a cellular solution with three cells obtained by the folding
of the (1,0) solution.

For the particular value $1/\nu=10.5$ (five poles) the different
possible solutions are shown on $\left[0,\pi\right]$ in Figure \ref{fig:shapes5,04,13,2}.
On the left, we have a monocoalescent (5,0) solution with five poles
at $0$. The middle solution of the figure is a (4,1) solution (4
poles at $x=0$, 1 pole at $x=\pi)$. Finally the solution on the
right is a (3,2) solution (3 poles at 0, 2 poles at $\pi$). For an
even value of the optimal number of poles (i.e. the value of $n$
in the interval $\left[2n-1,2n+1\right]$), the stable solutions will
include a solution symmetric on $\left[0,\pi\right]$, for instance
if $n=6$ we have the solutions (6,0) (5,1) (4,2) and the symmetric
(3,3) solution. In Figure \ref{fig:shapepole3,2} , we show on the
same figure the shape $(x,\phi(x))$ ($x$ is the horizontal direction)
of the (3,2) solution for $1/\nu=10.5$ with $x\in\left[0,\pi\right]$
(lower part of the figure, below the horizontal segment) and the corresponding
locations of poles in the complex plane (upper part of the figure,
above the horizontal segment). The poles are indicated by circles,
the segment is the real axis in pole space between $0$ and $2\pi$.
The important thing about this type of figure, which will be used
in the rest of the paper, is that a pole very close to the real axis
(i.e. very close to the horizontal segment in the upper part of the
figure) leads to a cusp in the front shape (in the lower part of the
figure), and that the $x$ value of the pole in the complex plane
is the same as the $x$ value in physical space of the cusp that appears;
in a diagram like Figure \ref{fig:shapepole3,2}, the cusp and the
corresponding pole are on the same vertical line . We will see later
however examples of poles far away from the real axis with no cusp
at the $x$ value of the pole. This effect results from a competition
between a new pole and the poles at zero which tend to prevent the
appearance of a new cusp. It is described in a simple way in Appendix
\ref{sec:How-far-must}.

An illustration of the stability of the (3,2) stationary solution
is given in Figure \ref{fig:stability3,2}. The initial condition
used in the Sivashinsky equation, with Neumann boundary conditions,
is exactly the (3,2) solution for $1/\nu=10.5$. In a simulation without
noise, the amplitude (maximum minus minimum of $\phi(x)$), would
simply stay constant with time, as the (3,2) solution is stable. In
order to complicate the convergence to the (3,2) solution, we apply
a noise (additive gaussian white noise added to the Sivashinsky equation,
amplitude $a=0.01$, see section \ref{sec:evolution-with-noise} for
other examples of simulation with noise, and other explanations) when
time < 10, and then continue the simulation without noise up to a
time of 500. The stability of the (3,2) stationary solution for Neumann
boundary conditions is illustrated by the fact that the shape returns
quickly to this solution (observe the fact that the final amplitude
is exactly the same as the initial one).

Of the different stable stationary solutions just described, the largest
basin of attraction (with initial conditions close to a flat flame
with some random perturbations) corresponds to the most symmetric
solution (i.e. the (3,2) solution for 5 poles) and the monocoalescent
solution ((5,0) in the previous case). It even seems, if one compares
both types of solutions, that the most symmetric solution has a larger
basin of attraction for low values of $1/\nu$ (in the case of five
poles for instance), and the monocoalescent one a larger basin for
large $1/\nu$. However, this result could be limited to this type
of initial conditions. Actually, in Section \ref{sec:evolution-with-noise},
it will be shown that in the presence of a moderate white noise added
to the Sivashinsky equation, the solution is much more often close
to the most symmetric bicoalescent solution with the optimal number
of poles than close to the corresponding monocoalescent solution.

Before closing this section, let us note the analogy of the bicoalescent
solutions found here with cellular solutions observed experimentally
in directional solidification \cite{jamgotchiantrivedibillia}. These
solutions, called doublets, look almost the same as the bicoalescent
solutions of this section. They are also stable for some range of
parameters. However, a major difference is that there is no instability
at large scale in directional solidification, and that as a result,
the structure with one small cusp, one large cusp can be repeated
a number of times in the overall doublet cellular structure. But in
both cases, flames (bicoalescent solutions) and directional solidification
(doublets), these type of stationary solutions are related to the
well-known phenomenon of tip-splitting of curved fronts \cite{pelceacademicpress88}.

\section{web of stationary solutions \label{sec:web-of-stationary}}

As most of the solutions of the previous section were not found by
Guidi and Marchetti, only some trivial, cellular solutions obtained
by folding, such as the (2,2) solution, we investigate in this section
higher values of $1/\nu$ than those used in their paper \cite{guidimarchetti}.
As in this paper, we plot the stationary solutions in a diagram giving
the amplitude (maximum minus minimum value of the solution) versus
$1/\nu$.

\begin{figure}

\caption{\label{fig:stable.eps}Stationary solutions: amplitude $\Delta\phi$
versus $1/\nu$ (light version with the monocoalescent solutions (n,0),
the cellular solutions (1,1,1,...) , and the stable bicoalescent solutions
(p,q))}

\includegraphics[%
  bb=41bp 53bp 686bp 613bp,
  scale=0.6]{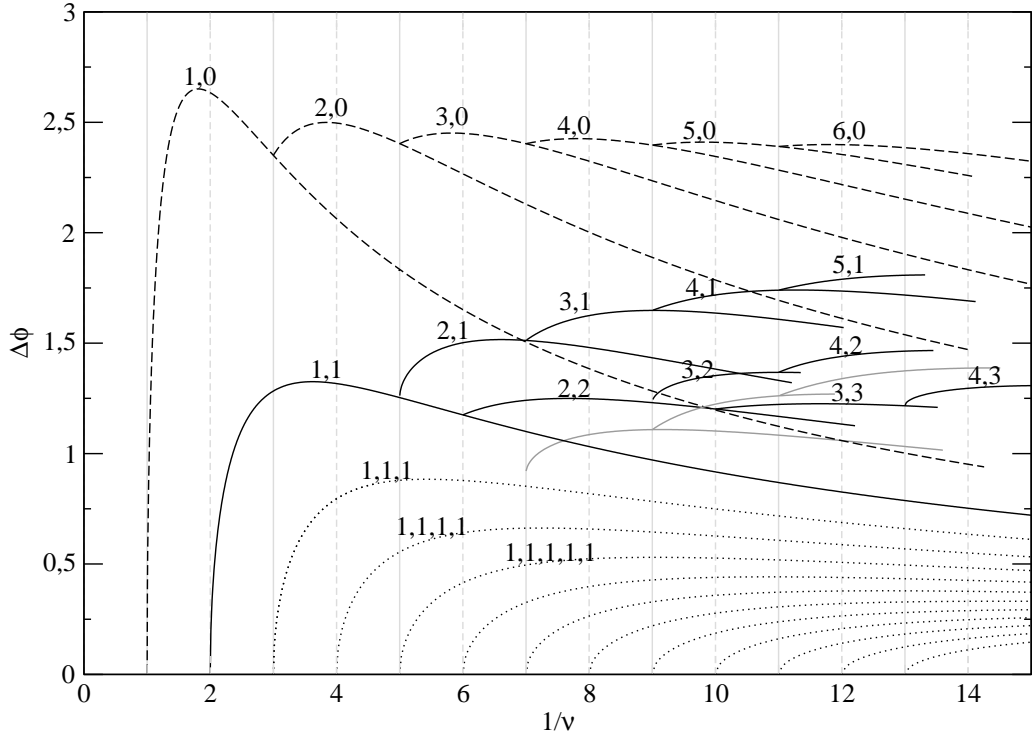}
\end{figure}

\begin{figure}[t]

\caption{\label{fig:all.eps}Stationary solutions: amplitude $\Delta\phi$
versus $1/\nu$ (complete version of the solutions obtained by the
author, including the interpolating solutions). }

\includegraphics[%
  bb=42bp 54bp 706bp 628bp,
  scale=0.6]{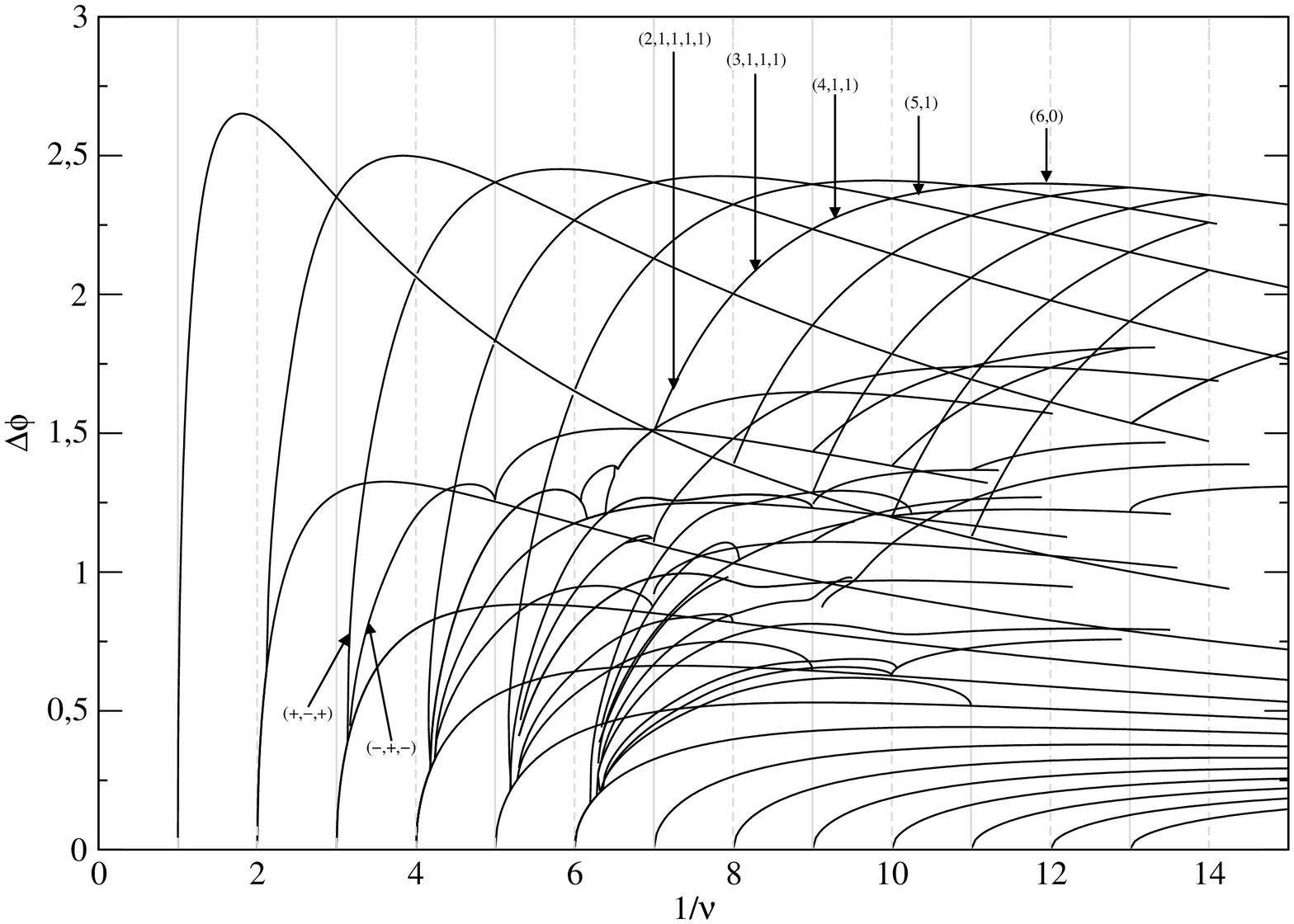}
\end{figure}

A light version of this diagram, with only the most important solutions,
particularly the bicoalescent solutions of the previous section, is
shown in Figure \ref{fig:stable.eps}. The complete version of this
diagram, with all the solutions obtained by the author, will be shown
in Figure \ref{fig:all.eps}. We have found it necessary to use two
figures, because the different solutions are so close in Figure \ref{fig:all.eps}
that it is difficult at first sight to recognize a particular bicoalescent
solution in this figure. We hope that a comparison between Figures
\ref{fig:stable.eps} and \ref{fig:all.eps} can help the reader understand
how the bicoalescent solutions of the previous section are interconnected
to the rest of the stationary solutions, particularly the cellular
ones. But the author knows, it is not an easy task for the reader,
so for the moment, we only start with the simplified version of the
diagram. To be more precise, we plot in Figure \ref{fig:stable.eps}
the basic solutions, i.e. the solutions with $n$ poles whose branch
exists in the interval $\left[2n-1,2n+1\right]$ of the parameter
$1/\nu$. In this interval, these type of solutions have thus the
optimal number of poles, a necessary condition for the solution to
be stable, as explained in Section \ref{sec:Stable-bicoalescent-solutions}.

In dashed lines in Figure \ref{fig:stable.eps} can be seen the monocoalescent
solutions $\left(n,0\right)$ which are created at $1/\nu=2n-1$ and
are stable in the periodic case until the next solution $\left(n+1,0\right)$
is created. From these solutions, by a process we call here folding
and which is defined in the previous section, the solutions (1,1)
(1,1,1) ... (dotted lines) are created, as well as the bicoalescent
(1,1) (2,2) (3,3) ... The non trivial bicoalescent solutions of Section
\ref{sec:Stable-bicoalescent-solutions} are created starting from
these symmetric bicoalescent solutions. The hierarchy (2,1) (3,1)
(4,1) (5,1)... is created starting from the (1,1) solution obtained
by folding. The hierarchy (3,2) (4,2) ... emerges from the (2,2) solution.
Finally, In Figure \ref{fig:stable.eps} the solution (4,3) (first
element of the hierarchy (5,3) (6,3) ...) is created from the (3,3)
solution, which means that one pole comes from infinity at a given
value of $1/\nu$ to create the solution.

All the solutions of the previous hierarchies are plotted as solid
black lines in Figure \ref{fig:stable.eps}. With the exception of
the folded symmetric solutions, all the other bicoalescent solutions
of this figure are stable when they are created, until a new solution
with one more pole appears. This behavior is exactly similar to the
monocoalescent solutions, the intervals of stability are also the
same. 

In solid gray lines in Figure \ref{fig:stable.eps} are plotted however
another hierarchy of solutions. This hierarchy contains solutions
of the type (2,1,1) (3,1,1) (4,1,1) apparently created exactly on
the same intervals as before. Of course this hierarchy only leads
to unstable solutions, in the periodic as well as the Neumann case.
It seems reasonable to suggest that as $1/\nu$ increases, an infinite
number of hierarchies will be created, each starting from a suitable
folded solution. The author actually suggests the following conjecture:
for each value $1/\nu$ of the control parameter with optimal number
of poles $n(\nu)$, all the multicoalescent stationary solutions having
the optimal number of poles, labelled $(n_{1},...,n_{p})$ for any
$p$ in the interval $1\leq p\leq n(\nu)$, with $\sum_{i=1}^{p}n_{i}=n(\nu)$,
do exist. 

Furthermore, as the amplitude of the solutions in these hierarchies
increases with $1/\nu$, it is extremely likely that solutions of
the $(n,1)$ hierarchy for instance, will soon become extremely close
to the corresponding monocoalescent solution $(n+1,0)$. And in the
Neumann case, all the bicoalescent hierarchies lead to stable stationary
solutions. A study of the time evolution of solutions of the Sivashinsky
equation will be reported in Section \ref{sec:evolution-with-noise}.

The previous argument suggests that there are many stationary solutions
of the Sivashinsky equation. However, as seen in Figure \ref{fig:all.eps},
Figure \ref{fig:stable.eps} was a very simplified version of the
diagram, with only the most important stationary solutions, which
were called basic solutions (see the explanation above), and form
a sort of skeleton of the entire structure of the solutions. We have
called this structure web of stationary solutions for obvious reasons,
all the solutions are interconnected, even the number of jumps necessary
to go from one solution to one another can probably be defined, reminiscent
of the hops from router to router on the internet. It is to be noted
that the other well-known Sivashinsky-type equation, the Kuramoto-Sivashinsky
equation, also admits a huge number of stationary solutions \cite{greenekim}.
The author does not even claim to have obtained in Figure \ref{fig:all.eps}
something comprehensive in the parameter space studied. The reader
is again warned that it is easier to look at both figures \ref{fig:stable.eps}
and \ref{fig:all.eps} at the same time, to locate first the basic
solutions that a particular interpolating solution connects.

The new solutions compared to Figure \ref{fig:stable.eps} are of
the interpolating type discussed by Guidi and Marchetti. We define
here these interpolating solutions (as opposed to basic solutions)
as solutions whose branch does not exist in the interval $\left[2n-1,2n+1\right]$
of the parameter $1/\nu$. Thus these solutions do not have the optimal
number of poles and cannot be stable (starting from such a solution,
a pole would come from infinity or disappear at infinity and a solution
with the optimal number of poles would be created). But in Figure
\ref{fig:all.eps}, it can be seen that these interpolating solutions
typically connect different basic solutions of the previous bifurcation
diagram (Figure \ref{fig:stable.eps}).

For instance, if one starts from the cellular solutions (1,1,1,...),
there exists interpolating solutions starting from this solution and
leading to all cellular solutions and the monocoalescent solutions
above. It must be noted that the precise values of $1/\nu$, where
these interpolating branches appear from the cellular solutions, were
calculated analytically in \cite{renardy}. In the simple case of
the (1,1,1) solution already studied by Guidi and Marchetti, it is
possible to move the poles vertically in the complex plane in two
different ways in order to have an initial guess of the position of
the poles on the interpolating branches (the Newton iteration leading
to the true values of the positions of the poles). Each interpolating
solution emanating from a cellular solution can be labelled by the
way the poles move along the interpolating branch compared to the
cellular solution. This type of pole movement along the interpolating
branch (at the beginning, where the branch is created) corresponds
exactly to the way the poles of the cellular solutions must be moved
in order to obtain an initial guess that will converge. So we have
the (+,-,+) solution: two poles are moved upward in the complex plane
(i.e. their imaginary part increases, while the real part is kept
constant), one downward compared to the (1,1,1) solution. This (+,-,+)
solution will interpolate, starting from the three cells solution,
all the monocoalescent solutions (1,0) (2,0) and (3,0) (this part
of the diagram will be described in more details later). We have also
the (-,+,-) solution, which, as seen in the figure, interpolates the
(1,1) solution (one pole going at infinity at this point). 

If we look at a much more complicated case, the five poles (1,1,1,1,1)
solution, it seems that in order to get the interpolating solutions,
we have to consider at least three levels of vertical movement of
the imaginary part of the poles, and for instance one interpolating
solution has been constructed by moving the third pole upward, the
first and fifth downward, the second and fourth somewhere in between.
Unfortunately, as shown in the case of the interpolating solutions
emanating from the six poles cellular (1,1,1,1,1,1) solution, the
author's capacities have been exceeded and neither the solution interpolating
(1,0) (2,0) (3,0) (4,0) (5,0) (6,0), neither the one interpolating
(1,1,1,1) have been found. Actually, although it is more or less obvious
that these solutions exist, the present author has been unable to
generate initial pole locations converging to these solutions (which
probably means that the author has not understood what type of perturbation
of the cellular solution leads to these two branches).

If the way the monocoalescent solutions are interpolated starting
from the cellular solutions is now considered, we prefer to start
now from the monocoalescent solution, for instance the (6,0) solution,
and decrease $1/\nu$. In Figure \ref{fig:stable.eps}, the monocoalescent
solutions were appearing suddenly apparently from nothing, for some
value of the control parameter. On the contrary, in Figure \ref{fig:all.eps},
precursors of the monocoalescent solution exist. So if the (6,0) solution
appears at $1/\nu=11$, what do we have exactly before ?

Actually, between $1/\nu=11$ and $1/\nu=10$, the precursor of (6,0)
is a bicoalescent (5,1) solution, with fives poles at zero, one at
$\pi$, however, the last one is very far from the real axis, and
does not lead to a cusp in the solution. This type of bicoalescent
solution, apart from the folded solutions like (2,2), were the only
ones obtained in Guidi and Marchetti (they have obtained actually
the (3,1) solution interpolating (4,0) and the (2,1) interpolating
(3,0)). They are unstable even for Neumann boundary conditions, because
they do not have the optimal number of poles corresponding to the
control parameter (the optimal number was defined in Section \ref{sec:Stable-bicoalescent-solutions}).

Between $1/\nu=10$ and $1/\nu=9$, the solution is no more bicoalescent,
but is instead a (4,1,1) solution. Then on $\left[8,9\right]$ we
have a (3,1,1,1) solution, on $\left[7,8\right]$ a (2,1,1,1,1) solution,
and as said before, we have not obtained the precursor close to the
six poles cellular solution. It is also possible to explain the previous
claim that the precursors of (6,0) interpolate all the monocoalescent
solutions with a number of poles less than 6. At $1/\nu=11$, one
of the six poles at zero goes to infinity, and reappears at $\pi$
to give a (5,1) solution. At $1/\nu=10$ , the pole at $\pi$ and
one of the poles at zero go to infinity, and reappear later to give
a (4,1,1) solution, and so on. The fascinating point is that although
all the precursors appear different, the curve of the amplitude of
all the precursors and of the final monocoalescent solution versus
the control parameter looks perfecly smooth. This, as well as the
overall structure of Figure \ref{fig:all.eps}, suggests that symmetries
less obvious than those leading to the folded solutions could be at
work in the Sivashinsky equation. 

\begin{figure}

\caption{\label{fig:frontcourbe}Different curved front solutions $\left(\phi(x)\right)$
with $x\in\left[0,\pi\right]$ for $1/\nu=10.5$. A constant has been
added to each solution in order to have the same spatial mean value
for all the solutions presented in this figure.}

\includegraphics[%
  bb=57bp 70bp 508bp 608bp,
  scale=0.5]{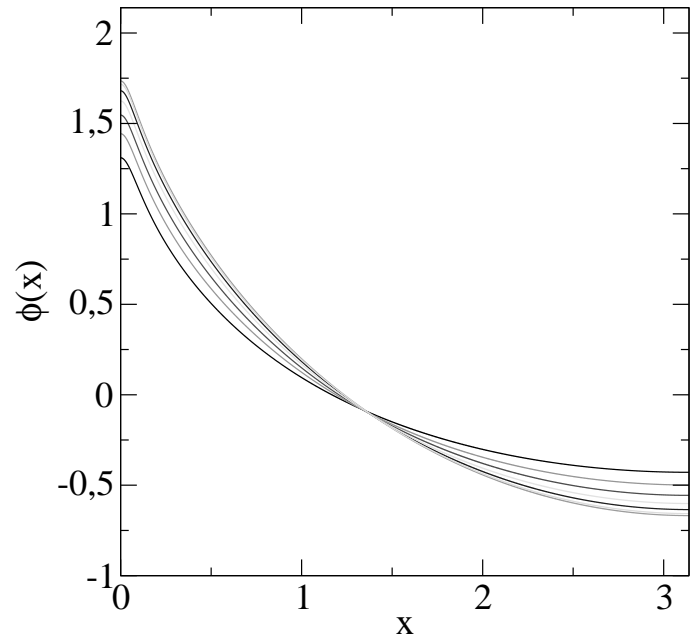}
\end{figure}

Let us look now at the shape of all these precursors in physical space.
We consider as before the case $\nu=10.5$ (optimal number of poles
: 5) . We show in Figure \ref{fig:frontcourbe} different curved flame
solutions. The one with the higher amplitude is the stable monocoalescent
(5,0) solution. Then we have, with smaller amplitude, a six poles
(5,1) solution interpolating (6,0). Then we have the four poles (4,0)
solution, a seven poles (4,1,1,1) solution interpolating (7,0), the
three poles (3,0) solution, and an eight poles (3,1,1,1,1,1) interpolating
(8,0). We have stopped there, as the next solutions in this list have
an amplitude very different from the original (5,0). The interesting
point is that in Figure \ref{fig:frontcourbe}, all these solutions,
which have a very different number of poles, look relatively similar,
like subsided versions the original monocoalescent solution, the first
ones being very close to (5,0) (and will be even closer with increasing
$1/\nu$). It seems that this is the way the Sivashinsky equation
is recovering a continuum of curved flame solutions in the limit $1/\nu\rightarrow\infty$,
something like the continuum of Ivantsov parabola of the related solidification
problem \cite{pelceacademicpress88}. From the simulations of Section
\ref{sec:evolution-with-noise}, it is not obvious at all that these
subsided unstable stationary solutions close to the monocoalescent
play any particular role in the dynamics, except perhaps by providing
ways to escape the stable monocoalescent solution during the transient
phase. We have shown in the successive Figures \ref{fig:Interpolating-(5,1)-solution},
\ref{fig:Interpolating-(4,1,1,1)-solution} and \ref{fig:Interpolating-(3,1,1,1,1,1)-solution}
the solutions and their poles for the non trivial cases (5,1), (4,1,1,1)
and (3,1,1,1,1,1) respectively. Once again, it highlights the fact
that the presence of poles is not equivalent to the presence of cusps,
sufficiently far from the real axis, and with other poles much closer,
some poles only lead to solutions with a weaker amplitude (see Appendix
\ref{sec:How-far-must}).

Now, if we take another look at the stable bicoalescent solutions
of Section \ref{sec:Stable-bicoalescent-solutions}, the same phenomenon
as for monocoalescent solutions has to be observed: the bicoalescent
solutions do not appear from nothing at a precise value of the parameters,
they have precursors, as seen in Figure \ref{fig:all.eps}. For instance,
we have produced precursors of the $\left(n,1\right)$ hierarchy,
which also look like subsided versions of the corresponding stable
bicoalescent solutions, and will also be closer to the original solution
as $1/\nu$ increases.

Overall, the bifurcation diagram going from cellular to curved flame
fronts with all the interpolating solutions of Figure \ref{fig:all.eps}
has a structure totally unexpected. In the Sivashinsky equation case,
most of the cellular solutions are unstable. However, the addition
of a sufficient amount of gravity (flames propagating downward) to
the Sivashinsky equation is known to stabilize these solutions and
to create a complex transition from cellular to curved fronts when
gravity is varied \cite{denetnonlinear}\cite{denetbonino}. It remains
to be seen if the structure of this transition has any relation with
Figure \ref{fig:all.eps}, which is likely, as a stable stationary
solution close to the bicoalescent solutions of the present paper
was found in \cite{denetbonino}. But searching for stationary solution
with gravity is much more difficult than with the Sivashinsky equation,
as no pole decomposition exists. The author takes this opportunity
to say that the instabilities of curved flames observed with a very
small gravity (and with periodic boundary conditions) in \cite{deneteurophys93}
would probably disappear with Neumann boundary conditions, as the
most violent instabilities of this paper are created by antisymmetric
modes.

\section{evolution with noise \label{sec:evolution-with-noise}}

\begin{figure}

\caption{\label{fig:avst_11.5_0.01_perio}Amplitude versus time with periodic
boundary conditions for $1/\nu=11.5$ and $a=0.01$}

\includegraphics[%
  bb=41bp 53bp 707bp 613bp,
  scale=0.5]{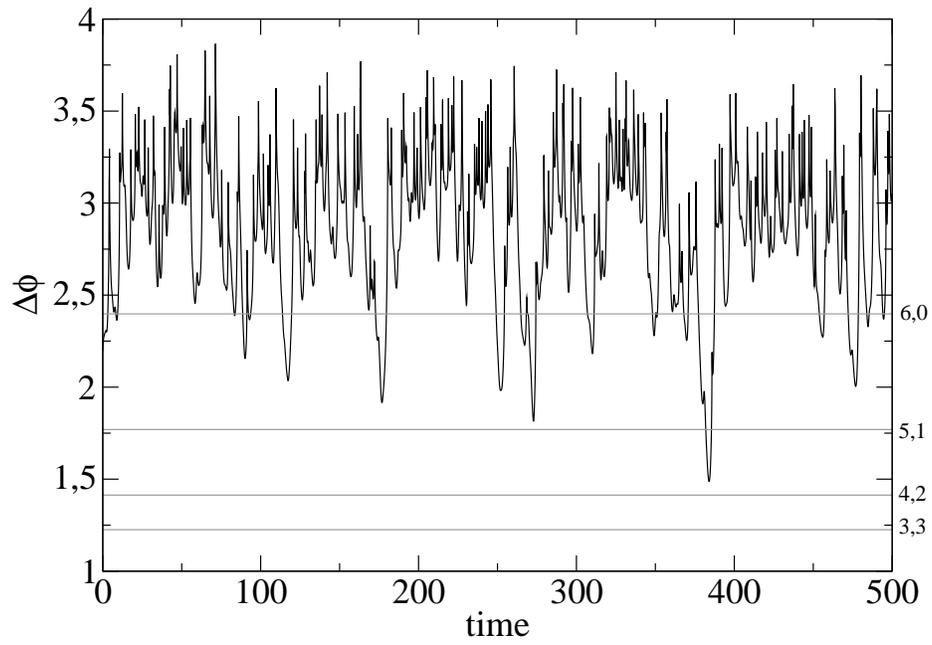}
\end{figure}

In Figure \ref{fig:avst_11.5_0.01_perio}, we start by showing a typical
time evolution with periodic boundary conditions, and a white noise
added to the right hand side of the Sivashinsky equation. This white
noise is gaussian, with zero mean value and deviation one, and we
multiply it by an amplitude $a$ . $a=0.01$ and $1/\nu=11.5$ (optimal
number of poles: six) in the simulations presented in this section,
with periodic and Neumann boundary conditions. In Figure \ref{fig:avst_11.5_0.01_perio}
is plotted, for periodic boundary conditions, the amplitude of the
front versus time, the initial condition being a five poles solution
which is not stationary for this value of the control parameter, and
leads to the initial transient. 

After this transient, the solution oscillates violently between low
and high values of the amplitude. The peak values correspond to curved
front solutions, with the poles being apparently almost monocoalescent,
but with an amplitude much higher than the monocoalescent (6,0) stationary
solution. The values of the amplitudes for the stationary solutions
(6,0) (5,1) (4,2) (3,3) are all indicated in the figure by gray lines,
so that the reader can compare. The low values correspond to shapes
with a new cusp formed in the flat part of the front. For the lowest
values of the amplitude, this new cusp leads almost to a bicoalescent
solution, but with again an amplitude which seems higher than the
(5,1) or (4,2) stationary solution. The solution never comes close
to the (3,3) solution, which on $\left[0,2\pi\right]$ is a two cells
solution. Furthermore, other low values of the amplitude correspond
to a cusp that develops without being exactly centered. Anyway, the
dynamics is dominated in the periodic case by antisymmetric perturbations.
Even if the new cusp formed by the perturbation is correctly centered
when it forms, it will ultimately move on one side, and will be swallowed
by the main cusp. This of course modifies the position of the main
cusp, and leads to the very high peak amplitudes observed. This antisymmetric
dynamics is forbidden for Neumann boundary conditions, so let us see
now what happens in this case.

\begin{figure}

\caption{\label{fig:avst_11.5_0.01_neumann}Amplitude versus time with Neumann
boundary conditions for $1/\nu=11.5$ and $a=0.01$}

\includegraphics[%
  bb=41bp 53bp 705bp 614bp,
  scale=0.5]{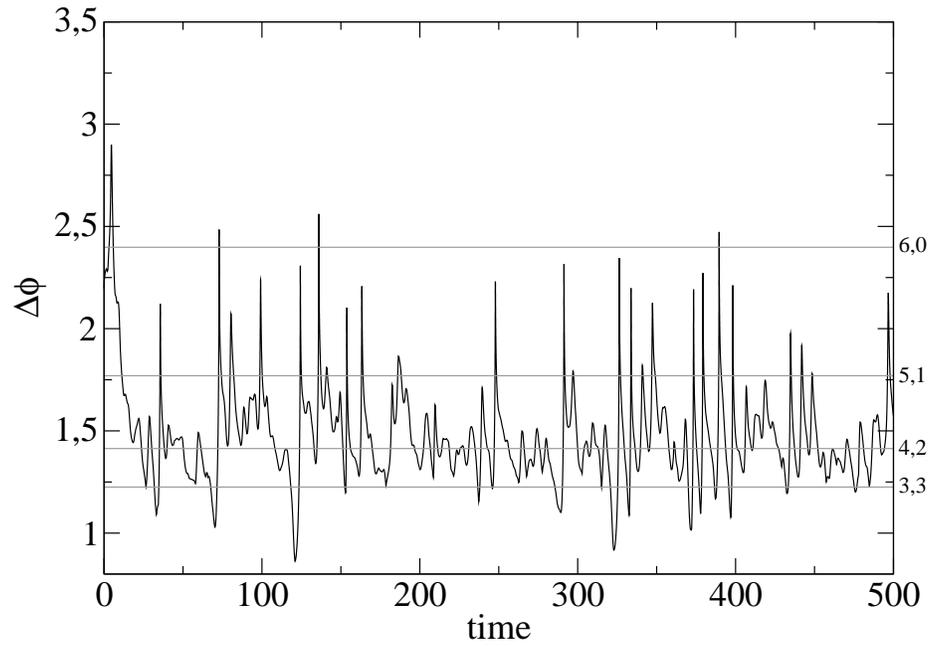}
\end{figure}

The situation is shown in Figure \ref{fig:avst_11.5_0.01_neumann},
for the same control parameter and noise amplitude as in the periodic
case. Before discussing this figure in detail, the overall impression
is that the signal obtained is much less turbulent. The different
stationary solutions for this value of the control parameter are also
indicated by gray lines.

The first point to note is that in this figure, except in the initial
transient, the front is never monocoalescent. Even for the peak values
obtained, where the amplitudes obtained sometimes seem close to the
(6,0) amplitude, we stress that all the solutions obtained at the
peak value are bicoalescent and not monocoalescent. On the contrary,
the solution seems often close to the different bicoalescent (3,3)
(4,2) and (5,1) solutions. We show in Figure \ref{fig:comparison_4,2_11.5neumann}
a comparison between the solution at time 410.555 in Figure \ref{fig:avst_11.5_0.01_neumann}
(dashed dotted line), where the amplitude has a local minimum very
close to the amplitude of the (4,2) solution, and the shape of the
(4,2) solution for $1/\nu=11.5$ (solid line). The agreement between
both solutions is excellent in this case. For very small values of
the noise amplitude (not shown here) the solution (with Neumann boundary
conditions) actually oscillates around the (3,3) solution, without
making jumps to any of the other stable stationary solutions. As the
noise used here is gaussian, it is not impossible however, that jumps
could occur as extremely rare events (for very small noise amplitudes),
and could be observed in very long simulations.

The value of the noise taken here $a=0.01$, although moderate, is
already sufficient to produce jumps in the amplitude, often actually
jumps between the bicoalescent steady solutions. The very low values
of the amplitude in Figure \ref{fig:avst_11.5_0.01_neumann} correspond
to shapes with three cusps in $\left[0,\pi\right]$, one on each boundary,
and one in the middle. For this value of the control parameter, the
middle cusp is always smaller than the cusps on the side. It is the
author's opinion that the lowest values of the amplitude correspond
to a shape close to an unstable stationary solution, which has not
been found in Figure \ref{fig:all.eps}. As the solution does not
need to be symmetric on $\left[0,\pi\right]$, the mechanism for the
disappearance of the middle cusp is relatively similar to the same
one on $\left[0,2\pi\right]$ in the periodic case, the middle cusp
moves on one side and is swallowed by one of the two main cusps. The
difference here with the periodic case is that the main cusp does
not move after having swallowed the small cusp and stays on the boundary.

After the low values of the amplitude comes a transient, where the
amplitude very quickly grows towards a peak value, which is a very
unstationary bicoalescent solution. Depending on the noise, the shape
will then often come back close to a stationary bicoalescent solution.
Finally, it seems that higher noise amplitudes or larger $1/\nu$
(the type of signal obtained is very sensitive to this last value)
lead to more turbulent curves of amplitude versus time with more jumps
and more time spent in the unstable low amplitudes solutions and the
very unstationary peaks.

To conclude this section, let us compare the behavior with Neumann
and periodic boundary conditions. For small $\nu$, the stable stationary
solutions are very sensitive to external noise in both cases. As is
well-known in the periodic case (and in this respect, the situation
is very similar with Neumann boundary conditions) , small perturbations
are continuously created on the front. But the difference lies in
the symmetries. In the periodic case, the stable stationary solutions
are the monocoalescent solution with the optimal number of poles,
and the continuum of its lateral translations, all neutrally stable
because of this symmetry. The noise keeps disturbing the monocoalescent
solution, but another solution of the continuum of monocoalescent
solutions (with the optimal number of poles) is also continuously
recreated. With Neumann boundary conditions, the stable solutions
are now the bicoalescent solutions with the optimal number of poles.
The perturbations created by the noise now serve to explore the different
stable bicoalescent solutions, causing jumps between two different
bicoalescent solutions. But with Neumann boundary conditions, all
stable solutions are not created equal, some are easier to destabilize
than the others. As seen previously for instance, the monocoalescent
solution is more sensitive to noise. As a result, during the time
evolution, the front will almost never be close to the monocoalescent
solution for small $\nu$ (which is just the opposite of the behavior
with periodic boundary conditions).

\section{conclusion \label{sec:conclusion}}

To summarize this paper, new bicoalescent solutions of the Sivashinsky
equation, stable in the Neumann case, have been obtained. They have
found their location in the incredible structure of the web of stationary
solutions. Simulations for moderate noise show that the evolution
is controlled by jumps between stationary solutions. The author would
like to insist here on the most important point of this paper: evolution
with periodic (controlled by antisymmetric perturbations) and Neumann
boundary conditions is very different. The Neumann boundary conditions
are more realistic, although in the presence of heat losses, the flame
is no more perpendicular to the wall (and is of course three dimensional).
Finally, it is likely that new analytical studies of the Sivashinsky
equation should be possible: even if the equation is now almost thirty
years old, many things remain to be explained.

\appendix

\section{How far must a pole be located from the real axis to create a new
cusp?\label{sec:How-far-must}}

In this appendix, we will try to explain in a very simplified way
that adding a new pole to a monocoalescent solution does not necessarily
create a new cusp if the isolated pole is located too far from the
real axis. Let us consider the following idealized situation: we have
a monocoalescent stationary solution with poles located at $0.$ A
new pole at $\pi$ is added to this solution, without moving any of
the other poles coalesced at $0.$ The front with the new pole is
no more stationary, but in this appendix, we try to answer the following
question: at which distance of the new pole to the real axis is a
new cusp created ? We call this distance $y_{c}$ and its value will
be measured numerically for different values of $1/\nu$, with an
optimal number of poles. In real situations the presence of the pole
at $\pi$ modifies the position of the poles at $0,$ particularly
the poles located far from the real axis. We neglect this effect as
we just want to have a reasonable order of magnitude of the value
of $y_{c}$ leading to a new cusp.

It turns out that the value of $y_{c}$ can be computed analytically
in the continuous approximation introduced by Thual Frisch and Hénon
\cite{thualfrischhenon}. Instead of summing on every pole located
at $0,$ this discrete sum is replaced by an integral, with a pole
density $\rho(y)$ ($y$ being the vertical coordinate in the complex
plane) given by (see \cite{thualfrischhenon} for a derivation):

\[
\rho(y)=\frac{1}{\pi^{2}\nu}\ln\left(\coth\frac{\left|y\right|}{4}\right)\]

The value of the slope of the front $\phi_{x}$corresponding to the
coalesced poles at 0 (in the continuous approximation) and to the
isolated pole at $\pi$ is given by:

\[
\phi_{x}\left(x\right)=-\nu P\int\rho\left(y\right)\cot\left(\frac{x-iy}{2}\right)dy-\nu\cot\left(\frac{x-\pi-iy_{1}}{2}\right)-\nu\cot\left(\frac{x-\pi+iy_{1}}{2}\right)\]

where $P$ denotes the principal value of an integral going from $-\infty$
to $+\infty$, the conjugated isolated poles being located at $\pi\pm iy_{1}$.
As a criterion for the appearance of a new cusp, we choose the natural
condition $\phi_{xx}\left(x=\pi\right)<0$. The value of $\phi_{xx}$
at this point is created by the competition between the coalesced
poles at $0$, which tend to prevent the creation of the new cusp,
and the isolated pole (and its complex conjugate) which has the opposite
effect. With the previous forms of the slope and the pole density,
we obtain:

\[
\phi_{xx}\left(x=\pi\right)=P\int\frac{1}{2\pi^{2}}\ln\left(\coth\left(\frac{\left|y\right|}{4}\right)\right)\frac{1}{\cosh^{2}\left(y/2\right)}dy-\frac{\nu}{\sinh^{2}\left(y_{1}/2\right)}\]

Integrating by parts, the antiderivative of the function under the
integral sign is 

\[
\frac{1}{\pi^{2}}\left(\ln\left(\coth\left(\frac{\left|u\right|}{2}\right)\right)\tanh\left(u\right)+2\arctan\left(\exp\left(u\right)\right)\right)\]

with $u=y/2$ , leading finally to

\[
\phi_{xx}\left(x=\pi\right)=\frac{1}{\pi}-\frac{\nu}{\sinh^{2}\left(y_{1}/2\right)}\]

In this formula, the term $1/\pi$ comes from the poles at $0$, the
other term from the isolated pole at $\pi.$ As said before, these
two terms have different signs. The condition $\phi_{xx}\left(x=\pi\right)=0$
finally leads to the value of $y_{1}=y_{c}$ corresponding to the
appearance of a cusp, which is:

\[
y_{c}=2\:\mathrm{arcsinh}\left(\sqrt{\pi\nu}\right)\]

\begin{figure}

\caption{\label{fig:yc}Comparison of the values of $y_{c}$ (maximum distance
to the real axis of a pole at $\pi$ to create a new cusp) obtained
numerically with a theoretical value obtained by using the continuous
approximation of Thual Frisch and Hénon}

\includegraphics[%
  bb=30bp 45bp 721bp 922bp,
  scale=0.5]{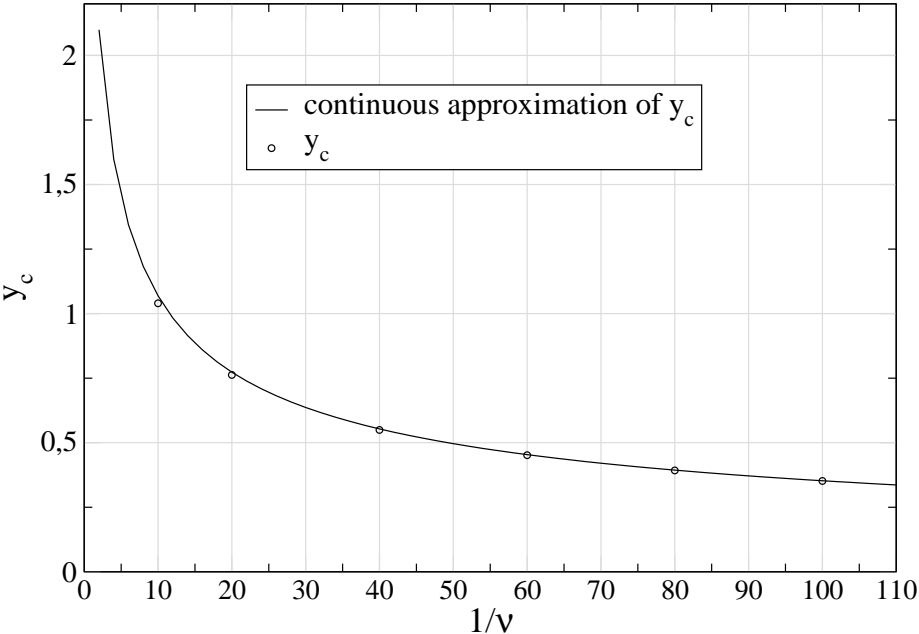}
\end{figure}

The cusp only appears if $y_{1}<y_{c}$. We now compare in Figure
\ref{fig:yc} this formula to the values of $y_{c}$ measured numerically
for $1/\nu=$10 (5 poles at $0$, one at $\pi$) , 20 (10+1 poles),
40, 60, 80, 100 (50+1 poles), each time with the optimal number of
poles coalesced at $0$ and one extra pole at $\pi$. The solid curve
is the previous formula obtained in the continuous approximation,
the circles are the values measured numerically. It can be seen that
the agreement is good. It is even more surprising if we think that
for $1/\nu=10$ we have only five poles at $0$ and the continuous
approximation for the second order derivative at $\pi$ works correctly,
the numerical point is just slightly below the theoretical curve.
Of course, this result is obtained in the framework of an illustrative
model where all the positions of the poles are kept fixed, but it
serves to justify the fact that in the presence of other poles, a
new pole at a different $x$ coordinate needs to be sufficiently close
to the real axis to create a new cusp.

\bibliographystyle{/usr/share/texmf/bibtex/bst/revtex4/apsrev}
\bibliography{./turb2d.reflib}

\begin{figure}[p]

\caption{\label{fig:shapes5,04,13,2}Flame shapes $\left(x,\phi(x)\right)$
with $x\in\left[0,\pi\right]$ of the (from left to right) (5,0) (4,1)
and (3,2) stationary solutions for $1/\nu=10.5$. All scales are the
same in the $x$ and $y$ direction}

\includegraphics[%
  scale=0.8]{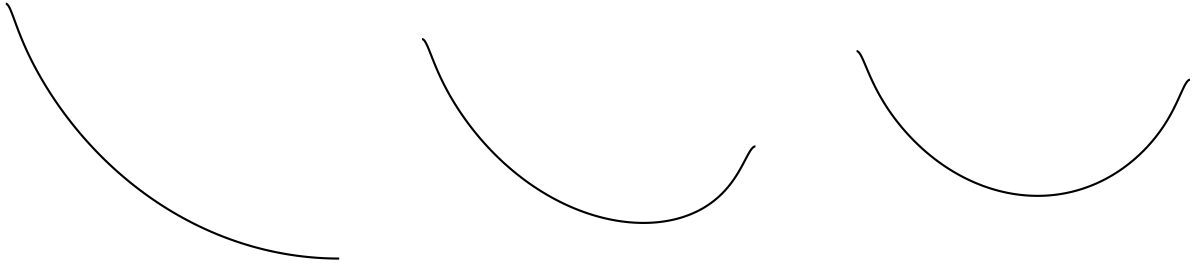}
\end{figure}

\begin{figure}[t]

\caption{\label{fig:shapepole3,2}Lower part of the figure (below the horizontal
segment) : flame shape $\left(x,\phi(x)\right)$ with $x\in\left[0,\pi\right]$
of the (3,2) stationary solution for $1/\nu=10.5$. Upper part of
the figure (above the horizontal segment) : corresponding pole locations
in the complex plane (the segment is the real axis in the complex
plane between $0$ and $2\pi$, the poles are indicated by circles).
All scales are the same in the $x$ and $y$ direction, both for the
flame shape and for the poles.}

\includegraphics[%
  scale=0.8]{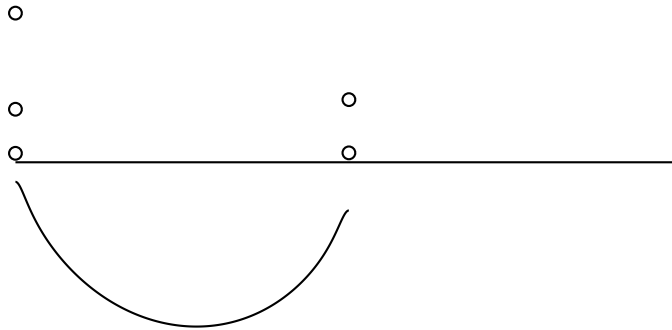}
\end{figure}

\begin{figure}

\caption{\label{fig:stability3,2} Amplitude versus time with Neumann boundary
conditions for $1/\nu=10.5$. The initial condition is the (3,2) stationary
solution. A gaussian white noise (amplitude $a=0.01$) is imposed
on this solution when time < 10, and is then suddenly stopped. The
solution goes back to the (3,2) solution for large times.}

\includegraphics[%
  bb=41bp 54bp 718bp 722bp,
  scale=0.4]{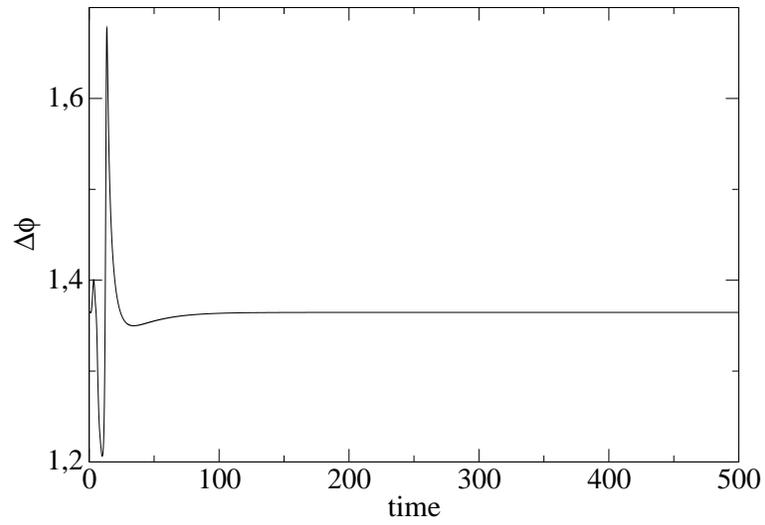}
\end{figure}

\begin{figure}

\caption{\label{fig:Interpolating-(5,1)-solution}Interpolating (5,1) solution
for $1/\nu=10.5$: lower part of the figure, flame shape, upper part
of the figure: pole locations (see Fig. \ref{fig:shapepole3,2} for
a more complete description of this kind of figure)}

\includegraphics[%
  scale=0.8]{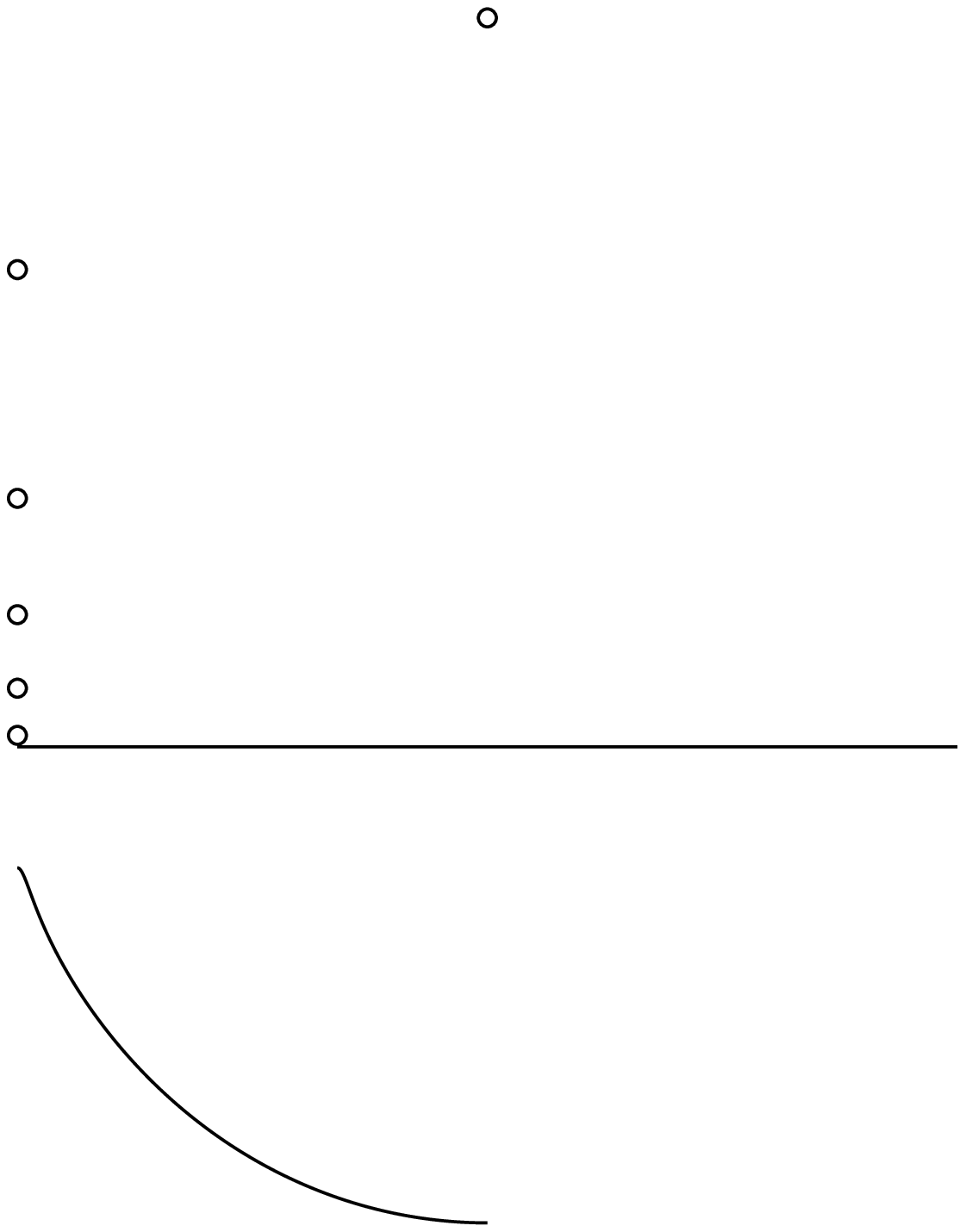}
\end{figure}

\begin{figure}

\caption{\label{fig:Interpolating-(4,1,1,1)-solution}Interpolating (4,1,1,1)
solution for $1/\nu=10.5$; lower part of the figure, flame shape,
upper part of the figure: pole locations (see Fig. \ref{fig:shapepole3,2}
for a more complete description of this kind of figure)}

\includegraphics[%
  scale=0.8]{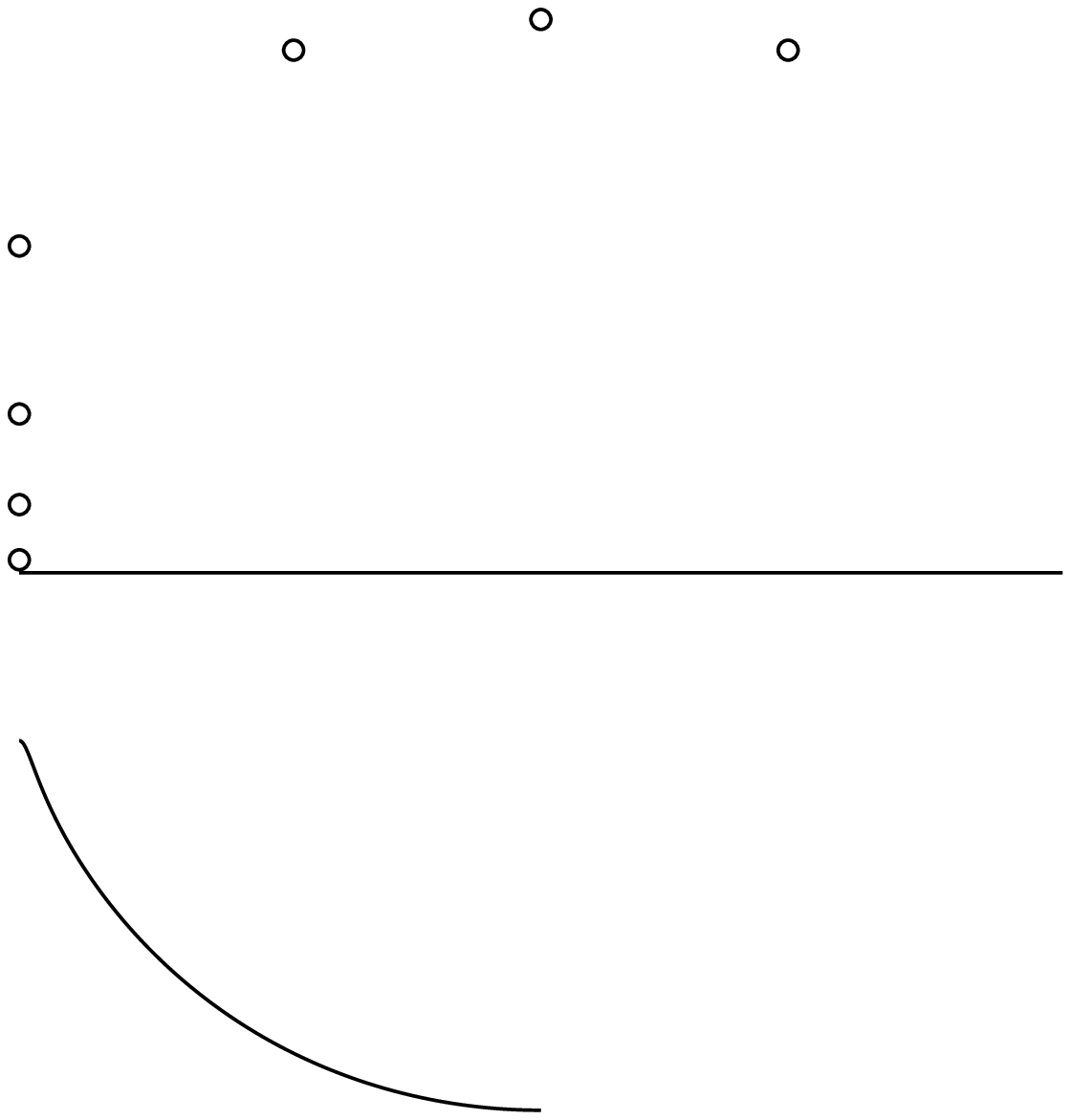}
\end{figure}

\begin{figure}

\caption{\label{fig:Interpolating-(3,1,1,1,1,1)-solution}Interpolating (3,1,1,1,1,1)
solution for $1/\nu=10.5$: lower part of the figure, flame shape,
upper part of the figure: pole locations (see Fig. \ref{fig:shapepole3,2}
for a more complete description of this kind of figure)}

\includegraphics[%
  scale=0.8]{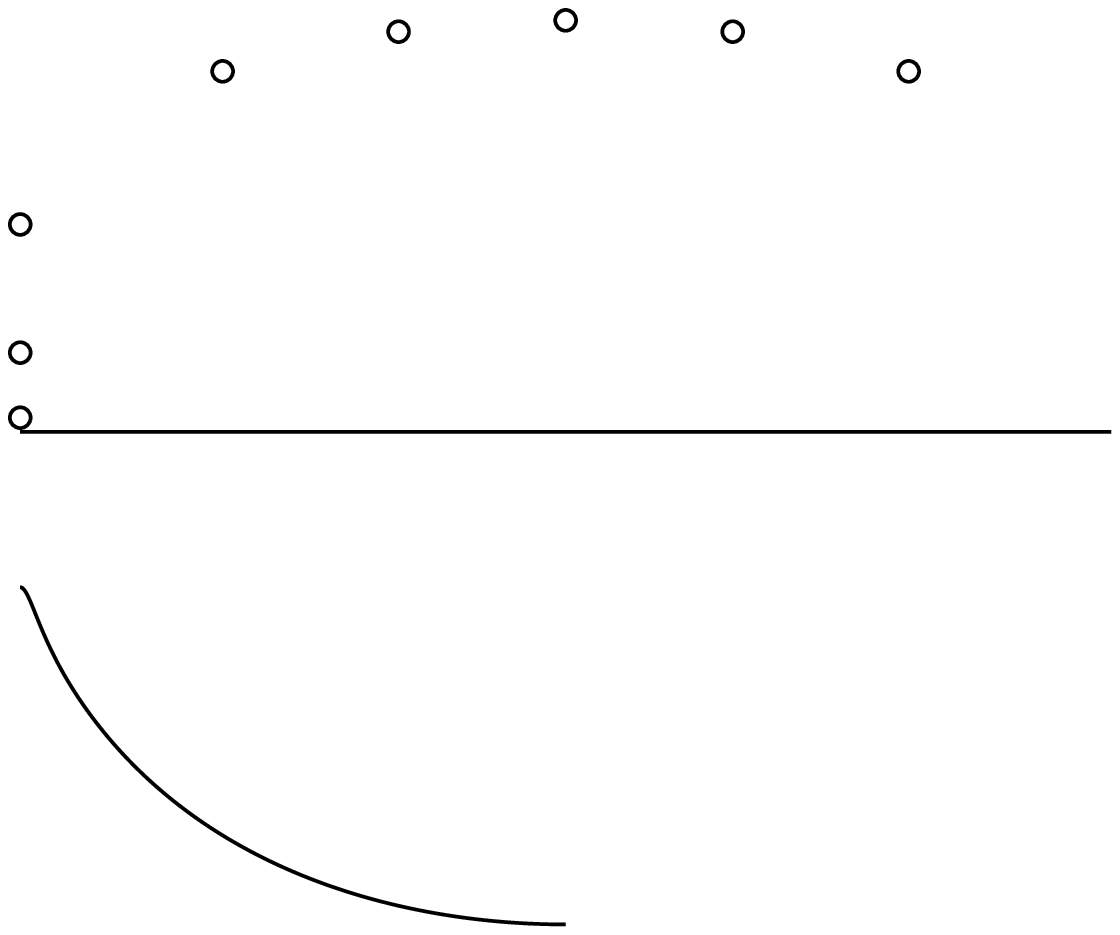}
\end{figure}

\begin{figure}

\caption{\label{fig:comparison_4,2_11.5neumann} Comparison of the solution
at time 410.555 (dashed dotted line) in Figure \ref{fig:avst_11.5_0.01_neumann}
(Neumann boundary conditions $1/\nu=11.5$ and $a=0.01$) with the
stationary (4,2) solution for $1/\nu=11.5$ (solid line)}

\includegraphics[%
  bb=37bp 59bp 622bp 622bp,
  scale=0.4]{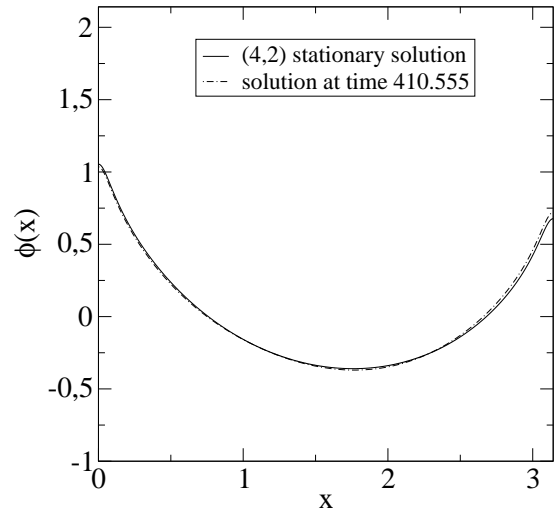}
\end{figure}

\end{document}